\documentclass[aps,pra,superscriptaddress,amsmath,amssymb,preprintnumbers,floatfix,showpacs,11pt]{revtex4}

\usepackage{amssymb}
\usepackage{epsfig}

\begin{document}
\title{ Linear atomic quantum  coupler }
\author{ Faisal A. A. El-Orany}
\email{el_orany@hotmail.com; faisal.orany@mimos.my }
 \affiliation{Department of Mathematics  and computer Science,
Faculty of Science, Suez Canal University 41522,
 Ismailia, Egypt;
 Cyberspace Security
Laboratory, MIMOS Berhad, Technology Park Malaysia, 57000 Kuala
Lumpur, Malaysia}

 \author{Wahiddin M. R. B.}
\email{mridza@mimos.my}
 \affiliation{ Cyberspace Security
Laboratory, MIMOS Berhad, Technology Park Malaysia, 57000 Kuala
Lumpur, Malaysia}

\date{\today}

\begin{abstract}
In this paper, we develop  the notion of the linear atomic quantum
coupler. This device consists of  two modes propagating into two
waveguides, each of them includes a localized and/or a trapped
atom. These waveguides are placed close enough to allow exchanging
energy between them via evanescent waves. Each mode interacts with
the atom in the same waveguide in the standard way, i.e. as the
Jaynes-Cummings model (JCM), and with the atom-mode in the second
waveguide via evanescent wave. We present the Hamiltonian for the
system and deduce the exact form for the wavefunction. We
investigate the atomic inversions and the second-order correlation
function. In contrast to the conventional linear coupler, the
atomic quantum coupler is able to generate nonclassical effects.
The atomic inversions can exhibit long revival-collapse phenomenon
as well as subsidiary revivals based on the competition among the
switching mechanisms in the system. Finally,   under certain
conditions, the system can yield  the results of the two-mode JCM.

\end{abstract}

  \maketitle
\section{Introduction}

 Quantum directional coupler is a device composed of two (or
more) waveguides placed close enough to allow exchanging energy
between them via evanescent waves \cite{jen1}. The rate of flow of
the exchanged energy  can be controlled by the device design and
the intensity of the input flux as well.
 The outgoing fields from the
coupler can be examined in the standard ways to observe the
nonclassical effects. Quite recently, this device has attracted
much attention in the framework of the optics communication and
quantum computing networks \cite{qcoup}, which require data
transmission and ultra-high-speed data processing \cite{EKer1}.
Furthermore, the directional coupler has been experimentally
implemented, e.g. in planar structures \cite{exp1}, dual optical
fibres \cite{exp2} and certain organic polymers \cite{exp3}. For
more details related to the quantum properties of the fields in
the directional couplers the reader can consult  the review paper
\cite{qu20} and the references therein.

The interaction between the radiation field and the matter (, i.e.
atom), namely, Jaynes-Cummings model (JCM) \cite{jay1}, is an
important topic in the quantum optics and quantum information
theories \cite{kni}. The simplest form of the  JCM is the
two-level atom interacting with the  single-mode of the radiation
field. The JCM is a rich source for the nonclassical effects, e.g.
the revival-collapse phenomenon (RCP) \cite{eber}, sub-Poissonian
statistics and squeezing \cite{fa}. Furthermore, the JCM has been
experimentally implemented by various means, e.g. one-atom mazer
\cite{remp}, the NMR refocusing \cite{meu}, a Rydberg atom in a
superconducting cavity \cite{sup}, the trapped ion \cite{vogele}
and  the micromaser \cite{micr}. Various extensions to the  JCM
have been reported including the two two-level atoms interacting
with the radiation field(s) \cite{tess,faisalob}.

The trapped atoms or molecules are promising systems for quantum
information processing and communications \cite{Nielsen}. They can
serve as convenient and robust quantum memories for photons,
providing thereby an interface between static and flying qubits
\cite{Lukin}. The subject of coupling cold atoms to the radiation
field sustained by an optical waveguide has already appeared in
various contexts. For example, hollow optical glass fibers were
used to guide atoms over long distances \cite{Noh}, especially,
employing red detuned light field filling out the hollow core
\cite{Letokhov,Renn}. Substrate based atom waveguide can also be
realized by using guided two-color evanescent light fields
\cite{Barnett}. Moreover, the coupling of atomic dipoles to the
evanescent field of tapered optical fibers has been demonstrated
in \cite{Vetsch,Nayak}. In this respect the  optical nanofibers
can manipulate and probe single-atom fluorescence.
 Moreover, it has been suggested that using a two-color evanescent light field around a
subwavelength-diameter fiber traps and guides atoms. The optical
fiber carries a red-detuned light and a blue-detuned light, with
both modes far from resonance. When both input light fields are
circularly polarized, a set of trapping minima of the total
potential in the transverse plane appears  as a ring around the
fiber. This design allows confinement of atoms to a cylindrical
shell around the fiber \cite{Kien}.  Additionally, it has been
shown that sub-wavelength diameter optical fibers can be used to
detect, spectroscopically investigate, and mechanically manipulate
extremely small samples of cold atoms. In particular, on
resonance, as little as two atoms on average, coupled to the
evanescent field surrounding the fiber, already absorbed 20  of
the total power transmitted through the fiber.  By optically
trapping one or more atoms around such fibers \cite{Dowling}, it
should become possible to deterministically couple the atoms to
the guided fiber mode and to even mediate a coupling between two
simultaneously trapped atoms \cite{Fam}. This leads to a number of
applications, e.g., in the context of quantum information
processing, high precision measurements,
 single-photon generation in optical fiber or
EIT-based parametric four-wave mixing \cite{horak}
 using a few atoms around optical
nanofibers.  Inspired by these  facts we develop here the notion
of the atomic quantum coupler (AQC), for which the interaction
mechanisms inside the waveguides and between the waveguides depend
on both the atomic and bosonic systems.   These mechanisms are
more complicated than those in the JCM, as we shall show shortly.
For the AQC we show that the atomic inversions can exhibit long
revival-collapse phenomenon as well as subsidiary-revival patterns
based on the switching mechanisms  in the system. Furthermore,
under certain conditions, the system can give  the results of the
two-mode JCM. Also, the system is able to generate nonclassical
effects. It is worth mentioning that  the inclusion of  one atom
in one of the ports of the non-linear coupler  has been considered
in \cite{abdalla}. Nevertheless, the solution of the equations of
motion there is obtained by the rotation of axes, which does not
give complete information on the system.

We restrict the study in this paper to the development of the
Hamiltonian model, its dynamical wavefunction and how does it
work. These issues are discussed in section II.  Additionally, in
section III, we study two quantities, namely, the atomic
inversions and the second-order correlation functions.

\section{Model formalism and its wavefunction}

In this section we describe the linear directional atomic quantum
coupler (AQC) and derive its wavefunction. Also we discuss some
basic differences between this device and the conventional
directional coupler \cite{qu20}. Thus it is reasonable to shed
some light on the linear directional coupler, which is described
by the following Hamiltonian \cite{qu20}:

\begin{equation}\label{ins1}
\frac{\hat{H}}{\hbar}=\sum_{j=1}^{2}\omega_j\hat{a}_j^{\dagger}\hat{a}_j+\lambda
(\hat{a}_1\hat{a}_2^{\dagger}+\hat{a}_1^{\dagger}\hat{a}_2),
\end{equation}
where $\hat{a}_{1}\quad(\hat{a}_{1}^{\dagger})$ and
$\hat{a}_{2}\quad(\hat{a} _{2}^{\dagger} $) are the annihilation
(creation) operators of the first and the second  modes in the
first and the second waveguides with the frequencies $\omega_{1}$
and $\omega_{2}$;  $\lambda$  is the  coupling constant between
the waveguides. Basically this device operates as a quantum
switcher since it can switch the nonclassical effects as well as
the intensities of the modes propagating inside one of the
waveguides to the other \cite{jans}. In other words, it  can not
generate nonclassical effects by itself.
 For some reason that will be clear
shortly, we   calculate the mean-photon numbers for the
Hamiltonian (\ref{ins1}) when the two modes are  in the states
$|\alpha,0\rangle$. Thus we arrive at:
\begin{equation}\label{ins2}
    \langle\hat{a}_1^{\dagger}(T)\hat{a}_1(T)\rangle=|\alpha|^2\cos^2(T),\quad
    \langle\hat{a}_2^{\dagger}(T)\hat{a}_2(T)\rangle=|\alpha|^2\sin^2(T),
\end{equation}
where $T=\lambda t$. These equations indicate strong switching
mechanism in the linear coupler, where the intensity $|\alpha|^2$
in the first waveguide  has been completely switched to the other
one. Moreover, the mean-photon numbers can not exhibit the  RCP.

 Now we are in a position
to  develop  the AQC, which is the main object of the paper.
 The atomic coupler consists of two  waveguides, each of which includes a localized and/or a trapped atom.
 The wavegudies are placed close enough to each other to allow interchanging
  energy between them. The two atoms (in the different waveguides) are located very adjacent to each other.
   In each waveguide one mode
 propagates along and interacts with the atom inside
   in a  standard way as the JCM.
 The atom-mode in each waveguide interacts with the other
one via the evanescent wave. The  fields exited from the coupler
can be examined as single or compound modes by means of homodyne
detection to observe the squeezing of vacuum fluctuations, or by
means of a set of photodetectors to measure photon antibunching
and sub-Poissonian photon statistics in the standard ways.
 The scheme for the AQC  is depicted in Fig. 1. From this figure  and in the framework of  the
rotating wave  approximation (RWA) the Hamiltonian describing the
AQC can be expressed as:
\begin{figure*}
\includegraphics{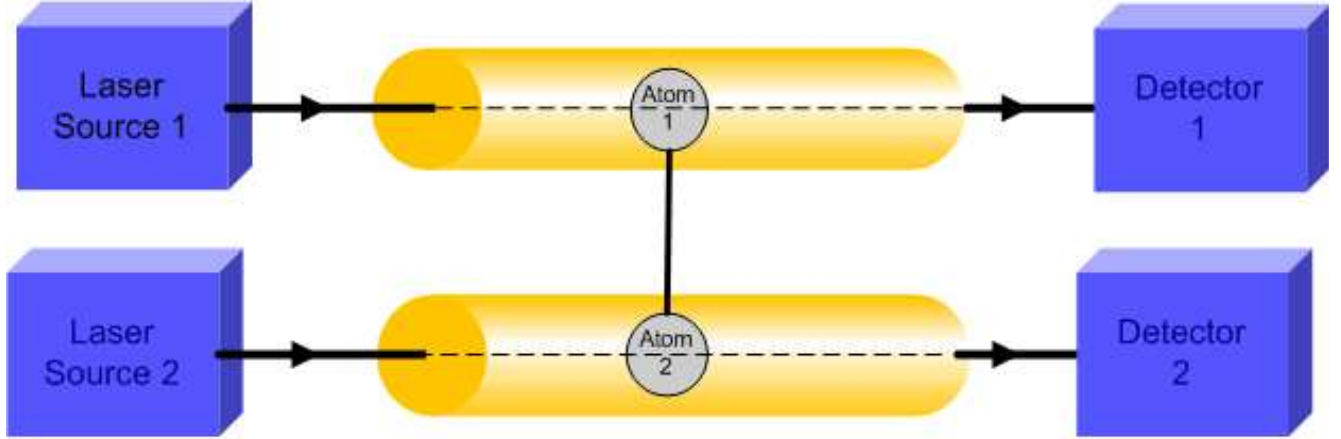}
\caption{\label{fig:wide} Scheme of realization of the Hamiltonian
(3). It is composed from two optical waveguides (yellow color).
The circles in these waveguides denote the localized and/or
trapped atoms. Mode 1 (2) pumped by, e.g., laser sources
propagates along the first (second) waveguide and interacts with
the first (second) atom via the coupling constant $\lambda_1\quad
(\lambda_2)$. The interaction between the first and the second
waveguide
 occurs  via the evanescent wave  with
the coupling constant $\lambda_{3}$.
 The outgoing fields from the
coupler can be measured  in the standard ways, e.g., using photon
detectors.}
\end{figure*}

\begin{widetext}
\begin{eqnarray}
\begin{array}{lr}
\frac{\hat{H}}{\hbar}=\hat{H}_0+\hat{H}_I,\\
\\
\hat{H}_0=
\sum\limits_{j=0}^{2}\omega_j\hat{a}_j^{\dagger}\hat{a}_j+
\frac{\omega_a}{2}(\hat{\sigma}_z^{(1)}+\hat{\sigma}_z^{(2)}),\quad
\hat{H}_I=\sum\limits_{j=1}^2 \lambda_j
(\hat{a}_j\hat{\sigma}_+^{(j)} + \hat{a}_j^{\dagger
}\hat{\sigma}_-^{(j)})+ \lambda_3 (\hat{a}_1\hat{a}_2^{\dagger
}\hat{\sigma}_+^{(1)}\hat{\sigma}_-^{(2)} +\hat{a}_1^{\dagger
}\hat{a}_2\hat{\sigma}_-^{(1)}\hat{\sigma}_+^{(2)}),
 \label{new1}
 \end{array}
\end{eqnarray}
\end{widetext}
where $\hat{H}_0$ and $\hat{H}_I$ are the free and the interaction
parts of the Hamiltonian, $\hat{\sigma}_\pm^{(j)}$ and
$\hat{\sigma}_z^{(j)}$ are the Pauli spin operators of the $j$th
atom ($j=1,2$); $\hat{a}_j\quad (\hat{a}_j^{\dagger})$ is the
annihilation (creation) operator of the $j$th-mode with the
frequency $\omega_j$  and $\omega_a$ is the atomic transition
frequency (we consider that the frequencies of the two atoms are
equal) and $\lambda_1\quad (\lambda_2)$ is the atom-field coupling
constant in the first (second) waveguide in the framework of the
JCM. The derivation of the JCM Hamiltonian is well known, e.g.
\cite{lo}. The interaction between the modes in the two waveguides
occurs through the evanescent wave with the coupling constant
$\lambda_3$. This term is the only one, which is conservative and
can execute switching between the two waveguides.  Thus  it plays
an essential role in the behavior of the AQC. We should stress
that the switching mechanism occurs through the two JCMs (in the
two waveguides) and can be obtained by applying the RWA in each
individual waveguide. In other words, the quantity  $\lambda_3
(\hat{a}_1\hat{\sigma}_-^{(1)}\hat{a}_2^{\dagger
}\hat{\sigma}_+^{(2)} +\hat{a}_1^{\dagger
}\hat{\sigma}_+^{(1)}\hat{a}_2\hat{\sigma}_-^{(2)})$ is
nonconservative  and hence it is cancelled out.
  Finally, the treatment of the switching mechanism in (\ref{new1}) is
related to the notion of coupler, however,  the existence of atoms
in the waveguides has been taken into account.  In (\ref{new1})
the treatment is considered only at the moment when the two fields
interacting with atoms in the waveguides. Also when we treat the
atoms (fields) classically the Hamiltonian (\ref{new1}) tends to
that of the linear directional coupler (two-atom interaction).

The interaction of two two-level atoms with the two modes has been
considered in the optical cavity earlier
\cite{faisalob,eberlyy,Federico}, however, in the sense different
from that presented above. For instance, as a sum of two separate
Jaynes-Cummings Hamiltonians  to investigate the entanglement
\cite{eberlyy} as well as the entanglement transfer
 from a bipartite continuous-variable (CV) system
to a pair of localized qubits \cite{Federico}.
 Also, the quantum
properties of the system  of  two two-level atoms interacting with
the two nondegenerate cavity modes when the atoms and the field
are initially in the atomic superposition states and the
pair-coherent state  has been investigated in \cite{faisalob}.

Next, we evaluate the wave function for the Hamiltonian
(\ref{new1}). We assume that the two modes and atoms are initially
prepared in the coherent states $|\alpha,\beta\rangle$ and in the
excited atomic states $|e_1,e_2\rangle$, respectively. For
resonance case $2\omega_a=\omega_1+\omega_2$ one can easily prove
that $[\hat{H}_0,\hat{H}_I]=0$. Under these conditions, the
dynamical wave function describing the system can be expressed as:

\begin{widetext}
\begin{eqnarray}
\begin{array}{lr}
\mid \Psi (t)\rangle =\sum\limits_{n,m=0}^{\infty }C_{n,m}\left[
X_{1}(t,n,m)\mid e_{1},e_{2},n,m\rangle +X_{2}(t,n,m)\mid
e_{1},g_{2},n,m+1\rangle \right. \\
\\
+\left. X_{3}(t,n,m)\mid g_{1},e_{2},n+1,m\rangle
+X_{4}(t,n,m)\mid g_{1},g_{2,}n+1,m+1\rangle
\right],\\
\\
C_{n,m}=\exp(-\frac{1}{2}|\alpha|^2-\frac{1}{2}|\beta|^2)\frac{\alpha^n\beta^m}{\sqrt{n!m!}},
\label{new3}
\end{array}
\end{eqnarray}
\end{widetext}
where $|g\rangle$ stands for  atomic ground state. From the
Schr\"{o}dinger equation we obtain the following system of
differential equations:
\begin{widetext}
\begin{eqnarray}
\begin{array}{lr}
  i\dot{X}_1(t,n,m) = \lambda_2\sqrt{m+1}X_2(t,n,m)+\lambda_1\sqrt{n+1}X_3(t,n,m), \\
 i\dot{X}_2(t,n,m) = \lambda_2\sqrt{m+1}X_1(t,n,m)+\lambda_3\sqrt{(n+1)(m+1)}X_3(t,n,m)
 +\lambda_1\sqrt{n+1}X_4(t,n,m), \\
 i\dot{X}_3(t,n,m) = \lambda_1\sqrt{n+1}X_1(t,n,m)+\lambda_3\sqrt{(n+1)(m+1)}X_2(t,n,m)
 +\lambda_2\sqrt{m+1}X_4(t,n,m), \\
  i\dot{X}_4(t,n,m) =
  \lambda_1\sqrt{n+1}X_2(t,n,m)+\lambda_2\sqrt{m+1}X_3(t,n,m),
  \label{secf1}
\end{array}
\end{eqnarray}
\end{widetext}
where the superscript  "$.$" means differentiation w.r.t. time. In
the following, we give only the details related to the solution of
the coefficient $X_1(t,n,m)$, where the others can be similarly
treated.  Differentiating the first and last equations in
(\ref{secf1}) and re-substitute by the others we obtain:
\begin{eqnarray}
\begin{array}{lr}
  (\hat{D}^2+A_{n,m})X_1(t,n,m) = -(i\lambda_3 c_2D+c_1)X_4(t,n,m), \\
  (\hat{D}^2+A_{n,m})X_4(t,n,m) = -(i\lambda_3 c_2D+c_1)X_1(t,n,m), \\
  \hat{D}=\frac{d}{dt}, A_{n,m}=\lambda_1^2(n+1)+\lambda_2^2(m+1),
  c_1=2\lambda_1\lambda_2\sqrt{(n+1)(m+1)},
  c_2=\lambda_3\sqrt{(n+1)(m+1)}.  \label{secf2}
\end{array}
\end{eqnarray}
From (\ref{secf2}) one can easily obtain:
\begin{equation}\label{adds}
    (\hat{D}^2+A_{n,m})^2X_1(t,n,m) = (i\lambda_3
    c_2D+c_1)^2X_1(t,n,m).
\end{equation}
This equation can be easily solved. By means of the initial
conditions stated above the exact  forms of the coefficients $X_j$
can be expressed as:
\begin{widetext}
\begin{eqnarray}
\begin{array}{lr}
 X_1(t,n,m)=\frac{1}{2}
\exp(i\frac{t}{2}c_2)\left[\cos(t\Omega_-)
-i\frac{c_2}{2\Omega_-}\sin(t\Omega_-)\right]
 +\frac{1}{2} \exp(-i\frac{t}{2}c_2)\left[\cos(t\Omega_+)
+i\frac{c_2}{2\Omega_+}\sin(t\Omega_+)\right],\\
\\
 X_2(t,n,m)=
\frac{-i\sqrt{m+1}}{2c_2^2
[A_{n,m}-4\frac{\lambda_1^2\lambda_2^2}{\lambda_3^2}] }\Bigl\{
 \exp(i\frac{t}{2}c_2)\left[(c_2^2-2c_1)\lambda_2^3(m+1)+(2A_{n,m}-c_1)\left(\lambda_2c_1-
\frac{\lambda_1\lambda_3}{2}(n+1)c_2\right)\right]\\
\\
\times
\frac{\sin(t\Omega_-)}{\Omega_-}+\exp(-i\frac{t}{2}c_2)\left[(c_2^2+2c_1)\lambda_2^3(m+1)
-(2A_{n,m}+c_1)\left(\lambda_2c_1-
\frac{\lambda_1\lambda_3}{2}(n+1)c_2\right)\right]\frac{\sin(t\Omega_+)}{\Omega_+}\Bigr\},
\\
\\
 X_3(t,n,m)=
\frac{-i\sqrt{n+1}}{2c_2^2
[A_{n,m}-4\frac{\lambda_1^2\lambda_2^2}{\lambda_3^2}] }\Bigl\{
 \exp(i\frac{t}{2}c_2)\left[(c_2^2-2c_1)\lambda_1^3(n+1)+(2A_{n,m}-c_1)\left(\lambda_1c_1-
\frac{\lambda_2\lambda_3}{2}(m+1)c_2\right)\right]\\
\\
\times
\frac{\sin(t\Omega_-)}{\Omega_-}+\exp(-i\frac{t}{2}c_2)\left[(c_2^2+2c_1)\lambda_1^3(n+1)
-(2A_{n,m}+c_1)\left(\lambda_1c_1-
\frac{\lambda_2\lambda_3}{2}(m+1)c_2\right)\right]\frac{\sin(t\Omega_+)}{\Omega_+}\Bigr\},
\\
\\
 X_4(t,n,m)=\frac{1}{2} \exp(i\frac{t}{2}c_2)\left[
- \cos(t\Omega_-) +i\frac{c_2}{2\Omega_-}\sin(t\Omega_-)\right]
 +\frac{1}{2} \exp(-i\frac{t}{2}c_2)\left[
\cos(t\Omega_+) +i\frac{c_2}{2\Omega_+}\sin(t\Omega_+)\right]
 , \label{ap1}
 \end{array}
\end{eqnarray}
\end{widetext}
 where
\begin{equation}\label{secf3}
\Omega_\pm=\frac{1}{2}\sqrt{\lambda_3^2(n+1)(m+1)+4(\lambda_1\sqrt{n+1}\pm\lambda_2\sqrt{m+1})^2}.
\end{equation}
It is obvious that the Rabi oscillation in the AQC is more
complicated than that of the JCM. From the solution (\ref{ap1})
different limits can be checked. For instance, when
$(\lambda_2,\lambda_3)\rightarrow (0,0)\quad(\lambda_3\rightarrow
0)$ the coefficients  (\ref{ap1}) reduce to those of the standard
JCM (two decoupled JCM \cite{eberlyy}). Moreover, when
$(\lambda_1,\lambda_2)\rightarrow (0,0)$ the system reduces to a
simple form, which is in a good correspondence with the
conventional coupler (\ref{ins1}). Nevertheless, the device, in
this case, is a rich source for the nonclassical effects. This
depends on the types of initial atomic states and can be explained
as follows: (i) The atoms are initially prepared in
$|e_1,e_2\rangle$. In this case the system reduces to the dark
state, where $\hat{H}_{int}|e_1,e_2\rangle=0$. These states do not
evolve in time. This property has been exploited in the quantum
clock synchronization \cite{synch}. (ii) The atoms are initially
prepared in  $|e_1,g_2\rangle$.  The dynamical state of the system
takes the form:

\begin{eqnarray}
\begin{array}{lr}
\mid \Psi (T)\rangle =\sum\limits_{n,m=0}^{\infty }C_{n,m}\left[
\cos[ T\sqrt{(n+1)(m+1)}]\mid e_{1},g_{2},n,m+1\rangle
\right.\\
\\
\left.-i\sin [T\sqrt{(n+1)(m+1)}]\mid g_{1},e_{2},n+1,m\rangle
\right], \label{dark1}
\end{array}
\end{eqnarray}
where $T=\lambda_3 t$. The expression (\ref{dark1}) reveals that
the behavior of the radiation fields   is typically that of the
two-mode single-atom JCM \cite{twjcm}. Finally, when the two atoms
are initially in the Bell state
$[|e_1,g_2\rangle+|g_1,e_2\rangle]/\sqrt{2}$ the wavefunction
takes the form:
\begin{widetext}
\begin{eqnarray}
\begin{array}{lr}
\mid \Psi (T)\rangle
=\frac{1}{\sqrt{2}}\sum\limits_{n,m=0}^{\infty }C_{n,m}\exp[-i
T\sqrt{(n+1)(m+1)}]\left[ \mid e_{1},g_{2},n,m+1\rangle +\mid
g_{1},e_{2},n+1,m\rangle \right]. \label{dark2}
\end{array}
\end{eqnarray}
\end{widetext}
 It is evident that the system exhibits atomic trapping, i.e.
$\langle\hat{\sigma}^{(1)}_z(T)\rangle=\langle\hat{\sigma}^{(2)}_z(T)
\rangle=0$. Furthermore, the system is able to generate
nonclassical effects, in particular, in the quantities, which
depend on the off-diagonal elements of the density matrix such as
squeezing (we have checked this fact).

\begin{figure*}
\includegraphics{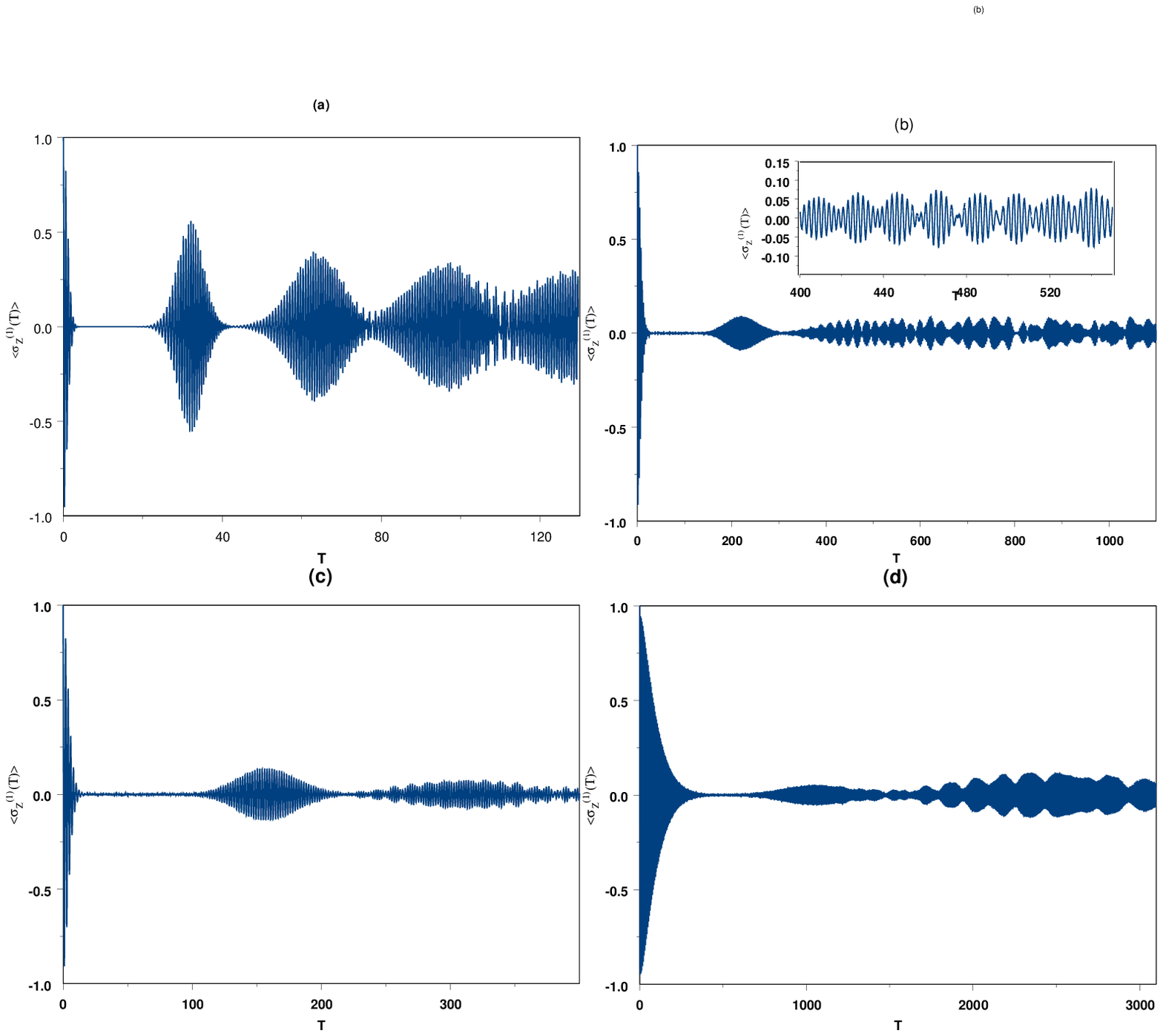}
\caption{\label{fig:wide2} Evolution of the
$\langle\hat{\sigma}^{(1)}_z(t)\rangle$
  against the interaction time $T=\lambda_1 t$ with $(\alpha, \beta)=(5,5)$ for
 $(\lambda_2,\lambda_3)=(1,0)$ (a), $(1,0.6)$ (b), $(2,3)$ (c) and $(1,1)$ (d).}
\end{figure*}

Now, we comment on the switching mechanism in the AQC. For the
sake of comparison, we substitute $\beta=0$ in relations
(\ref{new3})--(\ref{ap1}) and calculate the mean-photon numbers
as:

\begin{widetext}
\begin{eqnarray}
\begin{array}{lr}
\langle\hat{a}_1^{\dagger}(T)\hat{a}_1(T)\rangle=|\alpha|^2+\sum\limits_{n=0}^{\infty}
    |C_{n,0}|^2[|X_3(T,n,0)|^2+|X_4(T,n,0)|^2],\\
    \\
 \langle\hat{a}_2^{\dagger}(T)\hat{a}_2(T)\rangle=\sum\limits_{n=0}^{\infty}
    |C_{n,0}|^2[|X_2(T,n,0)|^2+|X_4(T,n,0)|^2], \label{ins3}
\end{array}
\end{eqnarray}
\end{widetext}
whre $T=t\lambda_1$. From these equations it is obvious that the
intensity of the mode in the first waveguide cannot be switched to
the other one. This is in a clear contrast with the linear
directional coupler (compare (\ref{ins2}) and (\ref{ins3})). This
behavior  is related to the nature of the atom-field interaction
mechanism, which is close to the classic Lee model of quantum
field theory. Moreover, this behavior is still valid even if the
interaction between the modes and the atoms  in the same waveguide
is neglected, i.e. $\lambda_1=\lambda_2=0$. In this case,
expressions (\ref{ins3}) exhibit the well-known RCP of the
standard JCM \cite{eber}.  The final remark, AQC is able to switch
the nonclassical effects from one waveguide to another based on
the values of the interaction parameters. This is remarkable from
(\ref{ins3}), where the mean-photon number in the second waveguide
$ \langle\hat{a}_2^{\dagger}(T)\hat{a}_2(T)\rangle$ can exhibit
RCP even though the second mode is initially  in vacuum state.
 On the
other hand, assume that the mode in the first waveguide is
initially prepared in the even coherent state, which can exhibit
squeezing, while the second mode is still in vacuum state.  In
this case,  the density matrix of the second mode takes the form:
\begin{equation}\label{density}
\hat{  \rho}_2=\sum\limits_{n=0}^{\infty}
    |C_{2n,0}|^2\Bigl\{[|X_1(T,2n,0)|^2+|X_3(T,2n,0)|^2]|0\rangle\langle 0|+
    [|X_2(T,2n,0)|^2+|X_4(T,2n,0)|^2]|1\rangle\langle1|\Bigr\},
\end{equation}
where $|C_{2n}|^2$  is the photon-number distribution of the even
coherent sate. From (\ref{density}), squeezing cannot be switched
to the second mode. Nevertheless, if the second mode is prepared
in the coherent state, it can exhibit squeezing. In this case, the
source of the nonclassical effects could be the switching
mechanism between the waveguides or the nature of the atom-field
interaction.

Now, we use above  relations  to investigate the atomic inversions
and second-order correlation functions in the following section.
For the sake of simplicity we consider $\alpha$ and $\beta$ to be
real.

\section{Atomic inversions and second-order correlation function}
Atomic inversion of the standard JCM is well known in quantum
optics by exhibiting RCP. The RCP has a nonclassical origin and
reflects the nature of the statistics of the radiation field. The
evolution of the atomic inversion has been realized via, e.g., the
one-atom mazer \cite{remp} and using technique similar to that of
the NMR refocusing \cite{meu}. In this section we investigate the
behavior of the AQC by studying the evolution of the atomic
inversions and the second-order correlation functions.
 As the system includes two atoms we have two types
of the atomic inversion, namely, single atomic inversion and total
atomic inversion
$\langle\hat{\sigma}_z(T)\rangle=\frac{1}{2}[\langle\hat{\sigma}^{(1)}_z(T)\rangle
+\langle\hat{\sigma}^{(2)}_z(T)\rangle]$. From (\ref{new3})  one
can obtain the following expressions:

\begin{figure*}
\includegraphics{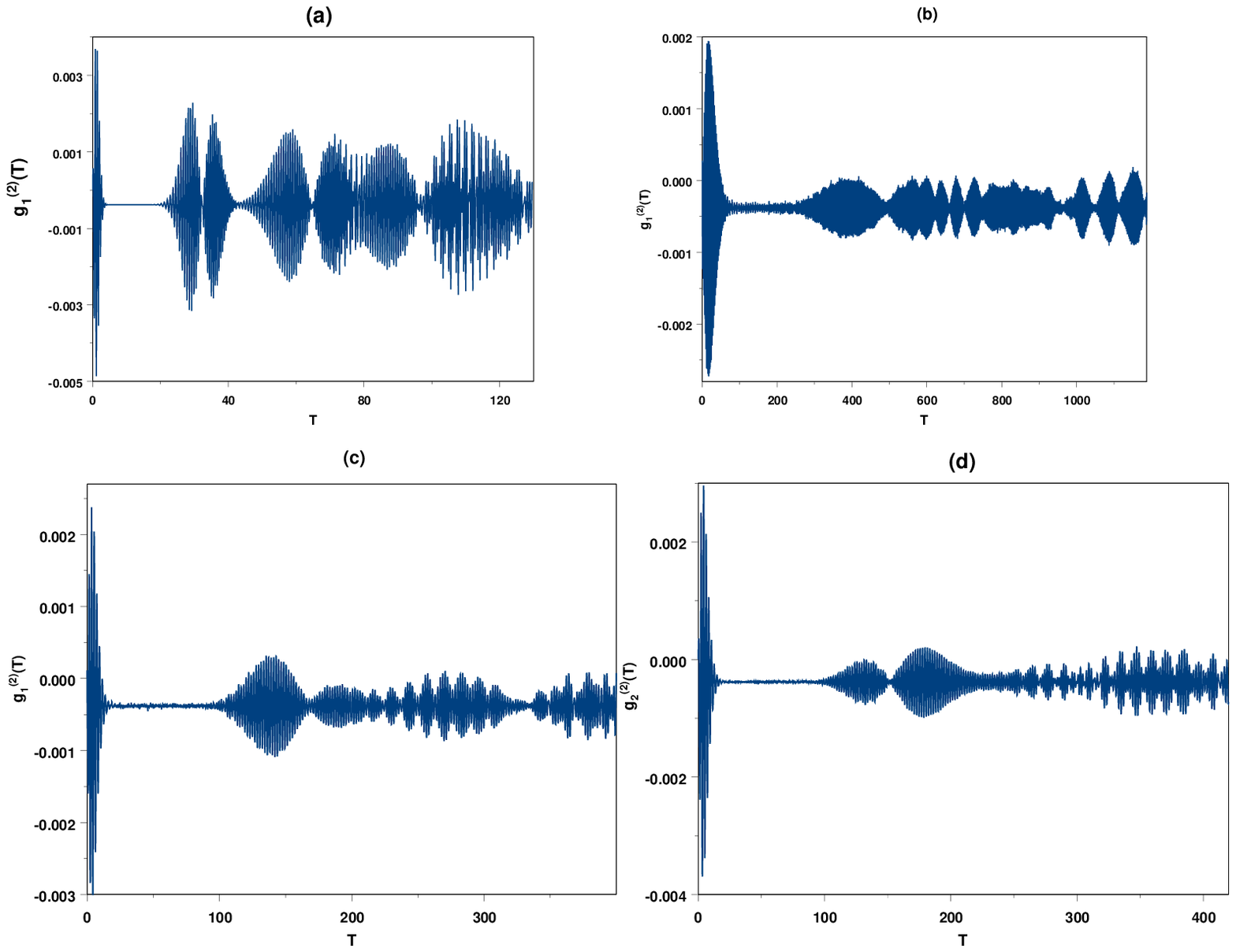}
\caption{\label{fig:wide3}  Evolution of the single-mode
second-order correlation function as indicated against the
interaction
 time $T=\lambda_1 t$ with $(\lambda_2,\lambda_3)=(1,0)$ (a), $(1,0.6)$ (b), $(2,3)$ (c)--(d).
 }
  \end{figure*}


\begin{widetext}
\begin{eqnarray}
\begin{array}{lr}
\langle\hat{\sigma}^{(1)}_z(T)\rangle =
\sum\limits_{n,m=0}^{\infty
}|C_{n,m}|^2[|X_{1}(T,n,m)|^2+|X_{2}(T,n,m)|^2
-|X_{3}(T,n,m)|^2-|X_{4}(T,n,m)|^2],\\
\\
\langle\hat{\sigma}^{(2)}_z(T)\rangle =
\sum\limits_{n,m=0}^{\infty
}|C_{n,m}|^2[|X_{1}(T,n,m)|^2-|X_{2}(T,n,m)|^2
+|X_{3}(T,n,m)|^2-|X_{4}(T,n,m)|^2],\\
\\
\langle\hat{\sigma}_z(T)\rangle = \sum\limits_{n,m=0}^{\infty
}|C_{n,m}|^2[|X_{1}(T,n,m)|^2-|X_{4}(T,n,m)|^2]. \label{new4}
\end{array}
\end{eqnarray}
\end{widetext}
  As we mentioned in the preceding section the conventional directional coupler
 cannot exhibit   RCP in the evolution of the mean-photon numbers.
 Nevertheless, the standard JCM
can exhibit RCP provided that the photon-number distribution  of
the initial field
 has a smooth envelope. Similar conclusion has been reported to
 the two-atom single-mode JCM
\cite{tess}. For the  AQC we have found when $\alpha=\beta$ and
$\lambda_j\neq 0$ the different types of the atomic inversions
(\ref{new4}) provide quite similar behaviors. It seems that  the
contributions of the coherence coefficients  $X_2, X_3$ are
comparable. Moreover, one can easily prove when $\lambda_3=0$ and
$\lambda_1=\lambda_2$ the atomic inversions reduce to  that of the
standard JCM (see Fig. 2(a)). It is worth reminding that for the
standard JCM the revival patterns occur in the atomic inversion
over certain period of the interaction time afterward they
interfere providing chaotic behavior. Additionally, the revival
time is connected with the amplitude $\alpha$ through the relation
$T_r=2\pi \sqrt{\bar{n}}\simeq 2\pi |\alpha|$ \cite{eber}. We
proceed, for $\lambda_1\neq\lambda_2$ they provide different forms
of the revival patterns. Here we restrict the attention to the
atomic inversion of the first atom (see Figs. 2(b)-(d) for the
given values of the interaction parameters).   We study three
cases based on the relationship between the strength of the
switching mechanisms in and between the waveguides, namely,
$\lambda_3<\lambda_j, \lambda_3=\lambda_j, \lambda_3>\lambda_j$.
Comparisons between Figs. 2(b)-(d) and Fig. 2(a) are instructive.
From Fig. 2(b) one can observe that the atomic inversion, after
the zero and first revival patterns, exhibits long series of the
subsidiary-revival patterns (see the inset in Fig. 2(b)).   This
behavior is completely different from that of the JCM. This
indicates that the nonclassical effects generated by this device
can sustain for an interaction time longer than that of the JCM.
It is worth mentioning that the subsidiary-revival patterns have
been observed for the JCM against the squeezed coherent state
\cite{eberr1}. This has been explained in relation to the
photon-number distribution of the initial states. More
illustratively, the photon-number distributions of the squeezed
states exhibit many peaks structure, each  of which gives its own
revival patterns in the evolution of the atomic inversion. These
patterns interfere with each other to produce these
subsidiary-revival patterns. Nevertheless, for the system under
consideration the occurrence of these patterns is related to the
switching mechanism between the waveguides (compare Figs. 2(a) and
(b)). This mechanism reflects itself in  very complicated Rabi
oscillations $\Omega_\pm$ as well as in the double summations in
the atomic inversions formulae  (\ref{new4}).
 Fig. 2(c) presents the case
when the coupling constants are different. It is obvious that the
RCP is still remarkable and the subsidiary revivals are smoothly
washed out compared to those in Fig. 2(b). Generally, we have
found when $\lambda_3\geq \lambda_1=\lambda_2$ the atomic
inversion exhibits  long RCP (see Fig. 2(d)). Above information
indicates that the switching mechanism between the waveguides
plays an important  role in the behavior of the AQC.
 Actually, we have found difficulties  in giving  mathematical
 treatment for the RCP presented by the AQC since the Rabi
oscillation  is rather complicated.

Now we draw the attention to  the second-order correlation
functions for the single-mode case, which is defined as:
\begin{equation}\label{pw3}
g_j^{(2)}(t)=\frac{\langle\hat{a}_j^{\dagger 2}(t)\hat{a}_j^{
2}(t)\rangle}{\langle
\hat{a}_j^{\dagger}(t)\hat{a}_j(t)\rangle^2}-1, \quad j=1,2,
\end{equation}
where $g_j^{(2)}(t)=0$ for Poissonian statistics (standard case),
$g_j^{(2)}(t)<0$ for sub-Poissonian statistics (nonclassical
effects) and $g_j^{(2)}(t)>0$ for super-Poissonian statistics
(classical effects). The second-order correlation function can be
measured by a set of two detectors, e.g. the standard Hanbury
Brown-Twiss coincidence arrangement. For the system under
consideration, this quantity  is plotted in Figs. 3 for the given
values of the interaction parameters. From these figures it is
obvious that the AQC is able to generate long-lived sub-Piossonian
effects, i.e. $g_1^{(2)}(t)<0$. Furthermore, the basic features of
the dynamics are still similar to those of the atomic inversion.
 Fig. 3(a)
presents the well-known  shape of the second-order function of the
standard JCM. When the switching mechanism between the waveguides
is involved the long RCP is dominant in the evolution of the
$g_j^{(2)}(t)$. Nevertheless, the shape of this phenomenon is
quite different from that in the corresponding atomic inversion
(compare Fig. 2(b) to Fig. 3(b)). For instance, the revival times
in the two quantities are different. Also, the number of the
subsidiary revivals in the atomic inversion is greater than that
in the corresponding $g_1^{(2)}(t)$.
  In contrast to the atomic
inversions, $g_1^{(2)}(t)$ and $g_2^{(2)}(t)$ can provide
different behavior for the same  values of the interaction
parameters. This fact can be realized by comparing Fig. 3(c) to
(d).

In conclusion, in this paper we have developed, for the first
time, the notion of the AQC. We have explained how does it work.
Also we have derived the exact solution for the equations of
motion. In contrast to the conventional coupler the AQC can
generate nonclassical effects. Nevertheless, the switching
mechanism in the former is more effective than that in the latter.
Furthermore, the behavior of the AQC is sensitive to the types of
the initial atomic states. We have shown that the system can give
the results of the two-mode JCM under certain conditions.
Additionally, we have  discussed the evolution of the  atomic
inversions and second-order correlation functions. These two
quantities can exhibit RCP, long RCP and long subsidiary-revival
patterns based on the values of the coupling constants.
Second-order correlation function can exhibit long-lived
nonclassical effects. From the information given in the
Introduction one can realize that the AQC is in the reach of the
current technology. Also it  may be of interest in the framework
of quantum information.

\section*{ Acknowledgement}

 The authors would like to thank Professor Jan Pe\v{r}ina for the interesting discussion.

\section*{References}


\begin{thebibliography}{200}
\bibitem{jen1} Jensen S M 1982 {\it
 IEEE J. Quant. Electron. QE} {\bf 18} 1580.
 \bibitem{qcoup}  Nikolopoulos G M 2008 {\it Phys. Rev. Lett.} {\bf 101}
200502.

\bibitem{EKer1}Ekert A and Jozsa R
1996 {\it Rev. Mod. Phys.} {\bf 68} 733; Lo H-K, Popescu S and
Spiller T 1998 "{\it Introduction to Quantum Computation and
Information}" (World Scientific: Singapore); Begie A, Braun D,
Tregenna B and Knight P L 2000 {\it Phys. Rev. Lett.} {\bf 85}
1762.

\bibitem{exp1}  Kamwa L P, Stitch J E, Mason N J and Roberts P N 1985
{\it Electron. Lett.} {\bf 21} 26; Jin R, Chuang C L, Gibbs H H,
Koch S W, Polky J N and Pubanz G A 1986 {\it Appl. Phys. Lett. }
{\bf 49} 110.
\bibitem{exp2}
 Gusovkii D D, Dianov E M, Mairer A A,
Neustreuev V B, Shklovskii E I and Scherbakov I A 1985 {\it Sov.
J. Quant. Electron.} {\bf 15} 1523.
\bibitem{exp3}
 Townsend P D, Baker G L, Shelburne J L III
and Etemad S 1989 {\it Proc. SPIE} {\bf 1147} 256.
\bibitem{qu20} Pe\v{r}ina Jr J and Pe\v {r}ina J 2000
{\it Progress in Optics} {\bf 41}, ed. E. Wolf (Amsterdam:
Elsevier), p. 361.



\bibitem{jay1} Jaynes E T and Cummings F W
 1963 Proc. IEEE  {\bf 51}  89.
\bibitem{kni}  Bose S,  Fuentes-Guridi I,  Knight P L and  Vedral V 2001 {\it Phys. Rev.
Lett.} {\bf 87}  050401.


\bibitem{eber} Eberly J H, Narozhny N B and  Sanchez-Mondragon J J
1980 {\it Phys. Rev. Lett.} {\bf 44} 1323;
 Narozhny N B,  Sanchez-Mondragon J J and Eberly J H 1981 {\it Phys. Rev. A}
 {\bf 23} 236;
Yoo H I,  Sanchez-Mondragon J J and Eberly J H 1981 {\it J. Phys.
A} {\bf 14} 1383; Yoo H I and Eberly J H 1985 {\it  Phys. Rep.}
 {\bf 118} 239.

\bibitem{fa} El-Orany F A A and Obada A-S 2003 {\it J. Opt. B: Quant. Semiclass. Opt.}
 {\bf 5} 60.

\bibitem{remp} Rempe G, Walther H and Klein N 1987 {\it Phys. Rev. Lett.}
{\bf 57} 353.
\bibitem{meu} Meunier T, Gleyzes S, Maioli P, Auffeves A, Nogues
G, Brune M, Raimond J M and Haroche S 2005 {\it Phys. Rev. Lett.}
{\bf 94} 010401.

\bibitem{sup} Yeazell J A, Mallalieu M and Stroud C R Jr 1990 {\it
Phys. Rev. Lett.} {\bf 64} 2007; Brune M, Schmidt-Kaler F, Maali
A, Dreyer J, Hagley E, Raimond J M and Haroche S 1996 {\it Phys.
Rev. Lett.} {\bf 76} 1800.
\bibitem{vogele} Vogel W and De Matos Filho R L 1995 {\it Phys. Rev.
A} {\bf 52} 4214.

\bibitem{micr} Meschede D, Walther H and M\"{u}ller 1985 {\it Phys.
Rev. Lett.} {\bf 54} 551;
 Walther H 1992 {\it Phys. Rep.} {\bf 219} 263.


\bibitem{tess} Tessier T E, Deutsch I H and Delgado A 2003 {\it Phys. Rev.
A} {\bf 68} 062316;    Faisal A A El-Orany 2006 {\it J. Phys. A:
Math. Gen.} {\bf 39} 3397; Faisal A A El-Orany 2006 {\it Phys.
Scripta} {\bf 74} 563.
\bibitem{faisalob} Faisal A A El-Orany, Obada A-S F, Abdelslama  M A and
Wahiddin M R B 2008 {\it J. Mod. Opt.} {\bf 55} 1649.

\bibitem{Nielsen}  Nielsen M and Chuang I 2000 "Quantum Computation and Quantum
Information" (Cambridge University Press, Cambridge); Lambropoulos
P and  Petrosyan D 2006 "Fundamentals of Quantum Optics and
Quantum Information" (Springer, Berlin).
\bibitem{Lukin} Lukin M D 2003 {\it Rev.
Mod. Phys.} {\bf 75} 457;  Petrosyan D 2005 {\it J. Opt. B} {\bf
7} S141.

\bibitem{Noh}
  Noh H-R and  Jhe W 2002 {\it Phys. Rep.} {\bf 372} 269.

\bibitem{Letokhov} Olshanii M
A,  Ovchinnikov Y B and  Letokhov V S 1993 {\it Opt. Comm.} {\bf
98} 77.
\bibitem{Renn}  Renn M J,  Montgomery D, Vdovin O, Anderson D Z, Wieman C E
and  Cornell E A 1995 {\it Phys. Rev. Lett.} {\bf 75} 3253.
\bibitem{Barnett}  Barnett A H, Smith S P, Olshanii M, Johnson K S,  Adams A W and  Prentiss
M 2000 {\it Phys. Rev. A} {\bf 61} 023608.
\bibitem{Vetsch}  Sague G, Vetsch E, Alt W, Meschede D and  Rauschenbeutel 2007 {\it Phys. Rev. Lett.} {\bf 99}  163602;
 Sague G, A Baade A and  Rauschenbeutel A 2008 {\it New. J. Phys.}
 {\bf 10} 113008.
\bibitem{Nayak} Nayak K P, Melentiev P N, Morinaga M,  Kien F L,  Balykin V I and  Hakuta K: quant-ph/0610136v1.

\bibitem{Kien} Kien F L,  Balykin V I and  Hakuta K 2004 {\it Phys. Rev. A} {\bf 70} 063403.

\bibitem{Dowling}
 Dowling J P and  Gea-Banacloche J 1996 {\it Adv. At. Mol. Opt. Phys.} {\bf 37} 1.

\bibitem{Fam} Kien F L,  Dutta Gupta S,  Nayak K P and
Hakuta K 2005 {\it Phys. Rev. A} {\bf 72} 063815.

\bibitem{horak}
  Horak P,  Domokos P and  Ritsch H 2003 {\it Europhys. Lett.} {\bf 61} 459.

 \bibitem{abdalla} Abdel-Aty M, Abdalla M S and Sanders  B C 2009
{\it Phys. Let. A} {\bf 373}  315.
\bibitem{jans}  Janszky J,  Sibilia C,   Bertolotti M and  Yushin Y
1988 {\it J. Mod. Opt.} {\bf 35} 1757.

\bibitem{lo}
  Louisell W H "Quantum Statistical Properties of
Radiation" (New York: Wiley 1973).



\bibitem{eberlyy} Y\"{o}naç M, Ting Y and  Eberly J H 2007 {\it J. Phys. B: At. Mol.
Opt. Phys.} {\bf 40} S45-S59.


\bibitem{Federico} Casagrande F,  Lulli A, and
Paris M G A 2007 {\it Phys. Rev. A} {\bf  75} 032336.

\bibitem{synch}  Jozsa R,  Abrams D S,  Dowling J P and  Williams C P 2000
{\it Phys. Rev. Lett.} {\bf 85} 2010.
\bibitem{twjcm}
 Gerry C C and  Eberly J H 1990 {\it Phys. Rev. A} {\bf 42} 6805;
Cardimona D A,  Kovanis V,  Sharma M P and  Gavrielides A 1991
{\it Phys. Rev. A} {\bf 43}  3710.

\bibitem{eberr1}
Satyanarayana M V, Rice P, Vyas R and Carmichael H J 1989  {\it J.
Opt. Soc. Am. B} {\bf 6} 228.

\end{thebibliography}
\end{document}